\documentclass[conference]{IEEEtran}
\usepackage[latin9]{inputenc}
\usepackage{color}
\usepackage{amsmath}
\usepackage{amssymb}
\usepackage{graphicx}
\usepackage{fancyhdr}
 
\makeatletter

\DeclareFontEncoding{LGR}{}{}
\DeclareRobustCommand{\greektext}{%
  \fontencoding{LGR}\selectfont\def\encodingdefault{LGR}}
\DeclareRobustCommand{\textgreek}[1]{\leavevmode{\greektext #1}}
\ProvideTextCommand{\~}{LGR}[1]{\char126#1}

\IEEEoverridecommandlockouts
\usepackage{cite}
\usepackage{amsfonts}\usepackage{algorithmic}
\usepackage{subfigure}
\usepackage{textcomp}
\usepackage{xcolor}
\def\BibTeX{{\rm B\kern-.05em{\sc i\kern-.025em b}\kern-.08em
    T\kern-.1667em\lower.7ex\hbox{E}\kern-.125emX}}
\makeatother

\begin{document}
\title{Frozen Mode in Three-Way Periodic Microstrip Coupled Waveguide}
\author{Mohamed Y. Nada, \textit{Member, IEEE}, Tarek Mealy, and Filippo Capolino,\textit{
Fellow, IEEE}}
\maketitle
\begin{abstract}
We demonstrate theoretically and experimentally that a periodic three-way
microstrip coupled waveguide exhibits a stationary inflection point
(SIP). The SIP is a third order exceptional point of degeneracy (EPD)
where three eigenmodes of the guiding system coalesce to form a stationary
frozen mode with zero group velocity. Here the frozen mode is shown
in a reciprocal waveguide; therefore three coupled waveguides (called
three-way waveguide) are required, supporting three modes in each
direction. We illustrate the occurrence of the frozen mode regime
by observing the dispersion diagram and by using the concept of coalescence
parameter that is a measure of the separation of the three coalescing
eigenvectors (polarization states). Results based on full-wave simulations
and measurements demonstrate the coalescence of three modes and how
this coalescence is not perfect due to the presence of losses in the
microstrip structure at 2 GHz. The coalescence parameter is used to
determine how close the three-way system is to the ideal frozen mode
condition.
\end{abstract}

\begin{IEEEkeywords}
Frozen modes, Coupled transmission line, Degeneracy, Stationary inflection
point, Exceptional point.

{\let\thefootnote\relax\footnotetext{This material is based upon work supported by the Air Force Office of Scientific Research award number FA9550-18-1-0355 and by the National Science Foundation under Grant Number  ECCS-171197. The authors would like to thank Dr. Ahmed Almutawa for his help with the fabrication and measurements of the structure. The authors are thankful to DS SIMULIA for providing CST Studio Suite that was instrumental in this study. \textit{(M. Nada and T. Mealy contributed equally to this work) (Corresponding author: Filippo Capolino)}}} 
{\let\thefootnote\relax\footnotetext{The authors are with the Department of Electrical Engineering and Computer Science, University of California at Irvine, Irvine, CA 92697 USA (e-mail: f.capolino@uci.edu).}} 
\end{IEEEkeywords}

\thispagestyle{fancy} 

\section{Introduction}

An exceptional point of degeneracy (EPD) is defined as the point in
a system parameter space at which two or more system eigenmodes coalesce
into a single degenerate mode \cite{figotin_oblique_2003}, and the
number of coalescing eigenmodes defines the order of the EPD. For
instance, the cut off frequency in any uniform waveguide is a second
order EPD resulting from the coalescence of two oppositely propagating
modes \cite{mealy2019degeneracy}, while the degenerate band edge
(DBE) is a fourth order EPD \cite{figotin_gigantic_2005,nada_theory_2017}.
EPDs are obtained also in systems with gain and losses exploiting
the concept of Parity-Time symmetry \cite{ruter_observation_2010,Heiss_PTsymmetry_2017},
however in this paper we focus on gainless waveguides.

This paper is concerned with a third order EPD in lossless and gainless
waveguides which is often referred to as the stationary inflection
point (SIP) or frozen mode regime. Such an EPD is obtained due to
the coalescence of three eigenmodes of which two are evanescent and
one is propagating to form \textit{a frozen mode} at which the group
velocity is zero. Indeed, the group velocity preserves its direction
for frequencies slightly smaller and higher than the SIP one, and
this makes it beneficial for various possible applications such as
amplifiers \cite{Yazdi_Third_2017} and Lasers \cite{Ramezani_Lasing_SIP_2014}.
The SIP was obtained in photonic crystals that support only four modes
(including both directions) through using magnetic materials to break
reciprocity \cite{figotin_nonreciprocal_2001,figotin_EM_SIP_2003,Mumcu_RF_MPC_2005,Sertel_frozen_2008,apaydin_experimental_2012}.
However, to obtain an SIP in reciprocal structures, at least three
coupled waveguides are required that allow three modes to exist in
each direction\cite{Li_Frozen_2017,Gan_Effects_2019}. The SIP has
been obtained in reciprocal optical waveguides by introducing periodicity
such as in periodic chain of coupled ring resonators that are side-coupled
to a straight waveguide \cite{nada_theory_2017}, three coupled waveguides
with periodic perforations \cite{Sertel_3way_SIP_2019}, coil resonators
\cite{Sumetsky_Fiber_2005,Scheuer_Optical_2011}, and also in \cite{gutman_slow_2012}
using coupled mode theory without referring to specific waveguide
design.

In this paper, we introduce a novel design of a three-waye waveguide
that exhibits an SIP at microwave frequencies shown in Fig. 1(a).
In Sec. \ref{sec:Frozen-Mode-in}, we provide the theoretical framework
and demonstrate the occurrence of the SIP in a lossless three-way
microstrip waveguide at 2 GHz using TL theory. In Sec. \ref{sec:Full-Wave-and-Experimental},
we demonstrate the existence of the SIP via full wave simulations
and experimentally, with both results being in very good agreement\textcolor{black}{.
We also show how the coalescing parameter is useful to assess the
coalescence of the three polarization state vectors.}

\section{\label{sec:Frozen-Mode-in}Frozen Mode in Lossless Three-Way Microstrip
Waveguide Design}

\begin{figure}
\centering \subfigure[]{\includegraphics[width=0.5\columnwidth]{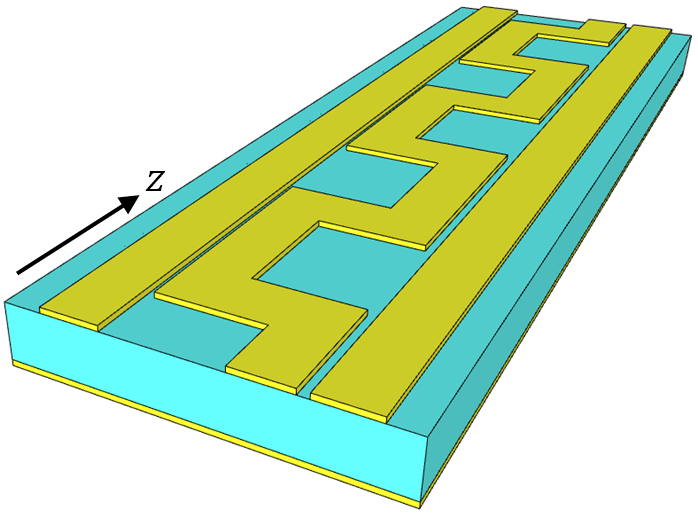}}
\subfigure[]{\includegraphics[width=0.25\columnwidth]{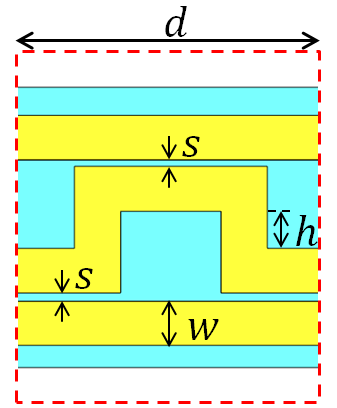}}

\centering \subfigure[]{\includegraphics[width=0.9\columnwidth]{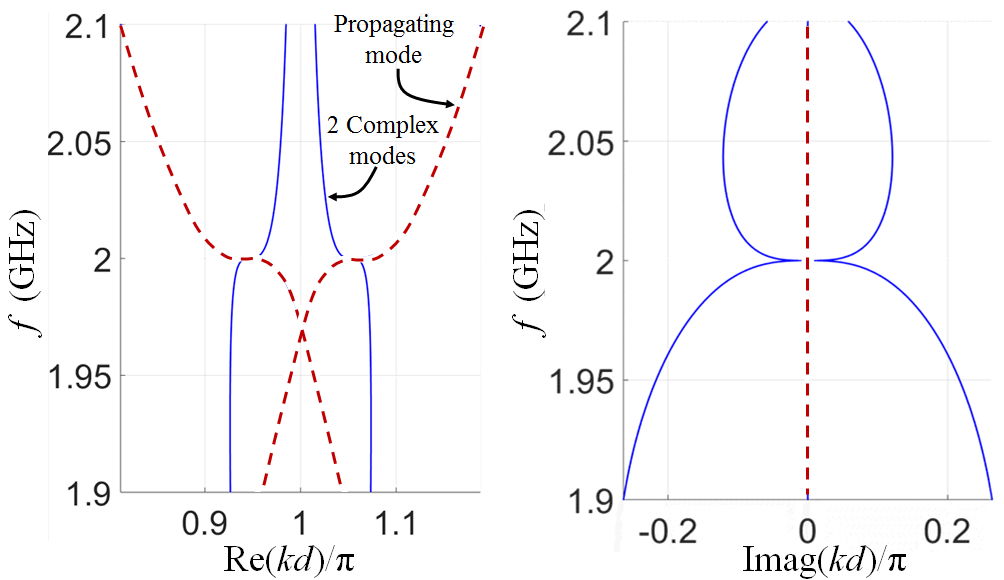}}

\caption{\label{fig:Proposed_structures}Three-way periodic microstrip waveguide
that exhibits an SIP (a frozen mode regime). (a) 3D perspective of
the geometry with copper lines over a grounded dielectric substrate
(metals are in yellow). (b) Top view of a unit-cell. The structure
can also be viewed as two straight TLs coupled through a serpentine
TL. (c) Floquet-Bloch complex-wavenumber dispersion diagram of eigenmodes
in the three-way periodic structure in (a). The dispersion diagram
shows the existence of an SIP at frequency $f=2$ GHz, where three
branches (one real and two complex) coalesce. The dispersion diagram
is shown for a lossless ideal microstrip, without considering conductor,
substrate, or radiation losses. The presence of losses perturbs the
exact occurrence of the SIP as shown in Fig. \ref{fig:Exp}. The dashed
red line represents the propagating modes with zero imaginary part.}
\end{figure}
The periodic three-way reciprocal microstrip, shown in Fig. \ref{fig:Proposed_structures}(a),
exhibits a frozen mode, i.e., an SIP in its $k$-$\omega$ dispersion
diagram, where $k$ is the Bloch-wavenumber and $\omega$ is the angular
frequency. The three-way waveguide comprises two uniform TLs that
are coupled through a third serpentine-shaped TL as shown in Fig.
\ref{fig:Proposed_structures}(a), with the unit cell shown in \ref{fig:Proposed_structures}(b).
We assume that all the TLs have the same width $w$ so that the three
individual TLs have a characteristic impedance of 50 Ohms (when uncoupled),
the distance between coupled lines is $s$, and the structure period
is $d$. To find the proper dimensions of the three-way microstrip
so that it exhibits an SIP at 2 GHz, we consider first a lossless
structure with substrate of a dielectric constant $\varepsilon_{r}=2.2$
and height $H=1.575$ mm. We determine the TL dimensions using the
quasi-static model in \cite{wheeler1965transmission} by imposing
the values of the desired characteristic impedance of 50 Ohms and
the dielectric permittivity of the grounded dielectric substrate.
The three-way microstrip supports six eigenmodes; three in each $z$-direction.
The proposed three-way microstrip shown here departs from the one
in \cite{apaydin_experimental_2012} because instead of breaking the
system reciprocity using nonreciprocal magnetic materials, we introduce
a third TL with periodic coupling to get the SIP while retaining the
system reciprocity. We define a state vector to describe the evolution
of the eigenmodes as $\psi=\left[\begin{array}{cccccc}
V_{1}, & I_{1}, & V_{2}, & I_{2}, & V_{3}, & I_{3}\end{array}\right]^{T}$, where $T$ denotes the transpose operation, and $V_{i}$ and $I_{i}$
with $i=1,2,3$ represent the voltage and the current in TL $i$,
respectively.

The evolution of the state vector of the periodic structure is described
by $\mathbf{\boldsymbol{\Psi}}(z_{2})=\mathbf{\underline{T}}(z_{2},z_{1})\mathbf{\boldsymbol{\Psi}}(z_{1})$,
where $\underline{\mathbf{T}}(z_{2},z_{1})$ is the $6\times6$ transfer
matrix (T-matrix) that translates the state vector from point $z_{1}$
to $z_{2}$. The spatial evolution of the state vector across a unit
cell of length $d$ is given by $\mathbf{\boldsymbol{\Psi}}(z+d)=\mathbf{\underline{T}}_{u}\mathbf{\boldsymbol{\Psi}}(z)$,
where the unit-cell T-matrix $\underline{\mathbf{T}}_{u}=\underline{\mathbf{T}}(z+d,z)$.
According to Floquet-Bloch theory, we look for periodic solutions
of the state vector as $e^{-jkd}$ where $k$ is the Floquet-Bloch
complex wavenumber, that satisfy $\mathbf{\boldsymbol{\Psi}}(z+d)=\lambda\mathbf{\boldsymbol{\Psi}}(z)$,
with $\lambda\equiv e^{-jkd}$. The eigenvalue problem is then formulated
as

\begin{equation}
\mathbf{\underline{T}}_{u}\mathbf{\boldsymbol{\Psi}}(z)=\lambda\mathbf{\boldsymbol{\Psi}}(z),\label{eq:Eig_prob}
\end{equation}
where the eigenvalues $\lambda_{n}\equiv e^{-jk_{n}d}$, with $n=1,2,\ldots,6$
are obtained by solving the dispersion characteristic equation $D(k,\omega)\equiv det[\mathbf{\underline{T}}_{u}-\lambda\mathbf{\underline{1}}]$,
with $\underline{1}$ being the $6\times6$ identity matrix. Due to
reciprocity, the determinant of $\underline{\mathbf{T}}_{u}$ is always
equal to unity \cite{EASWARAN_Relationship_1993} which implies that
the eigenvalues solutions of the characteristic equation must appear
in reciprocal pairs. This means that if $k$ is a Floquet-Bloch wavenumber
solution, then $-k$ is also a solution, i.e., the six \textit{k}-solutions
must come in positive-negative pairs. Away from an EPD, the unit-cell
T-matrix $\underline{\mathbf{T}}_{u}$ is diagonalizable hence it
has six eigenvectors $\boldsymbol{\Psi}_{n}(z)$ associated with six
eigenvalues $\lambda_{n}$. However at the EPD, the T-matrix $\underline{\mathbf{T}}_{u}$
is not diagonalizable and is similar to a matrix containing two $3\times3$
Jordan Blocks as shown in \cite{nada_theory_2017}. Since the waveguide
has 6 eigenvalues, at the SIP frequency we have only two reciprocal
eigenvalues $\lambda_{SIP}$ and $1/\lambda_{SIP}$, and each of them
has algebraic multiplicity 3 and geometrical multiplicity 1. Hence,
the six Floquet-Bloch wavenumber solutions are $k_{SIP}$ and $-k_{SIP}$,
and each one is repeated three times. The eigenvectors associated
with degenerate eigenvalues are generalized eigenvectors obtained
from $\left(\mathbf{\underline{T}}_{u}-\lambda_{SIP}\mathbf{\underline{1}}\right)^{p}\boldsymbol{\Psi}_{p}=\mathbf{\underline{0}},\quad p=1,2,3$,
where $3$ is the order of the SIP degeneracy, i.e., the number of
coalescing eigenvectors. Note that $\boldsymbol{\Psi}_{1}$ is the
regular eigenvector associated with the degenerate eigenvalue $\lambda_{SIP}$.

The SIP is a mathematical concept that is never met perfectly in reality
because losses and fabrication tolerances would inhibit the perfect
coalescence. However, a system can operate in the vicinity of the
frozen mode regime while retaining the unique physical properties
of the three-mode degeneracy. In order to assess how close a system
is to an EPD, we use the coalescence parameter concept developed in
\cite{abdelshafy_exceptional_2018} , where the authors referred to
it as figure of merit or hyperdistance. To better evaluate the coalescence
between three eigenvectors, we normalize the eigenvector terms so
that they all have the same unit as $\hat{\Psi}_{n}=diag(1,Z_{0},1,Z_{0},1,Z_{0})\Psi_{n}$,
with $n=1,2,\ldots,6$, and $Z_{0}$ is a normalization impedance
that is here considered to be equal to the TLs characteristic impedance.

The coalescence of the three eigenvetcors $\hat{\Psi}_{1}$, $\hat{\Psi}_{2}$
and $\hat{\Psi}_{3}$ associated to the three positive wavenumbers
in (\ref{eq:Eig_prob}) is measured via the coalescence parameter

\begin{equation}
\begin{array}{cc}
D_{H}=\dfrac{1}{3}\sum\limits _{\underset{n>m}{m=1,n=2}}^{3}\left|\sin\left(\theta_{mn}\right)\right|, & \cos\left(\theta_{mn}\right)=\dfrac{\mathit{\left|\left\langle \mathrm{\hat{\Psi}}_{m},\mathrm{\hat{\Psi}}_{n}\right\rangle \right|}}{\left\Vert \hat{\Psi}_{m}\right\Vert \thinspace\left\Vert \hat{\Psi}_{n}\right\Vert }\end{array}\label{eq:Hyperdistance}
\end{equation}
where $\theta_{mn}$ is the angle between the two six-dimensional
complex vectors $\hat{\Psi}_{m}$ and $\hat{\Psi}_{n}$, and it is
defined via the inner product $\left\langle \hat{\Psi}_{m},\hat{\Psi}_{n}\right\rangle =\hat{\Psi}_{m}^{\dagger}\hat{\Psi}_{n}$,
with the dagger symbol $\dagger$ representing the complex conjugate
transpose operation, and $\left\Vert \hat{\Psi}_{m}\right\Vert $
and $\left\Vert \hat{\Psi}_{m}\right\Vert $ denote their norms. The
parameter $D_{H}$ is always positive and less than one, and $D_{H}=0$
indicates the perfect coalescence of the three eigenvectors, i.e.,
the system experiences an SIP.

We build the transfer matrix of the unit-cell in Fig. \ref{fig:Proposed_structures}(b)
using transmission line analytic formulas by diving the waveguide
into sections of uniform single/coupled transmission lines and we
tuned the dimensions of the unit cell to minimize the three eignvectors
coalescence parameter $D_{H}$, where the eigenvecctors are calculated
from the eigenvalue problem (\ref{eq:Eig_prob}). The optimized lossless
unit cell has dimensions $w=5.09$ mm, $s=0.5$ mm, $h=3$ mm, and
a period $d=55$ mm, and it exhibits an SIP at 2 GHz as shown in Fig.
\ref{fig:Proposed_structures}(c), without using magnetic materials
to break the system reciprocity.

At the SIP, the dispersion relation is approximated as $\omega-\omega_{SIP}\approx\eta(k-k_{SIP})^{3}$,
where $\omega_{SIP}$ is the angular frequency at which the three
modes coalesce, $\eta$ is a constant that describes the flatness
of the SIP. The group velocity and its derivative are zero at the
SIP, i.e., $\partial\omega/\partial k=\partial^{2}\omega/\partial k^{2}=0$,
whereas the second derivative of the group velocity is non zero, i.e.,
$\partial^{3}\omega/\partial k^{3}=6\eta\neq0$.

\section{\label{sec:Full-Wave-and-Experimental} Experimental Verification
of SIP and Full-Wave Simulations}

\begin{figure}
\centering \subfigure[]{\includegraphics[width=0.68\columnwidth,height=0.3\columnwidth,keepaspectratio]{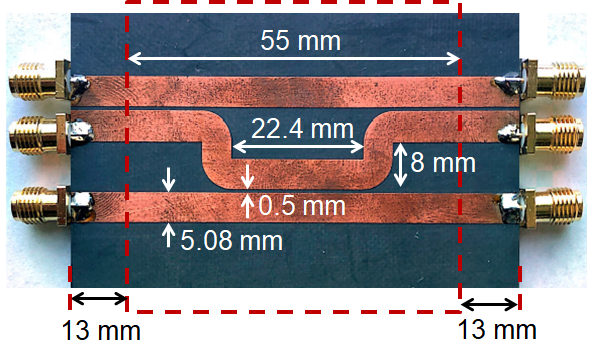}}\subfigure[]{\includegraphics[width=0.28\columnwidth,height=0.3\columnwidth,keepaspectratio]{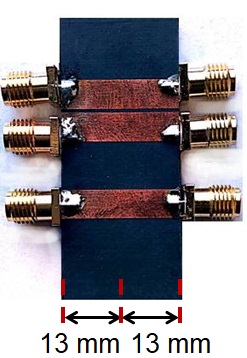}}

\subfigure[]{\includegraphics[width=0.9\columnwidth]{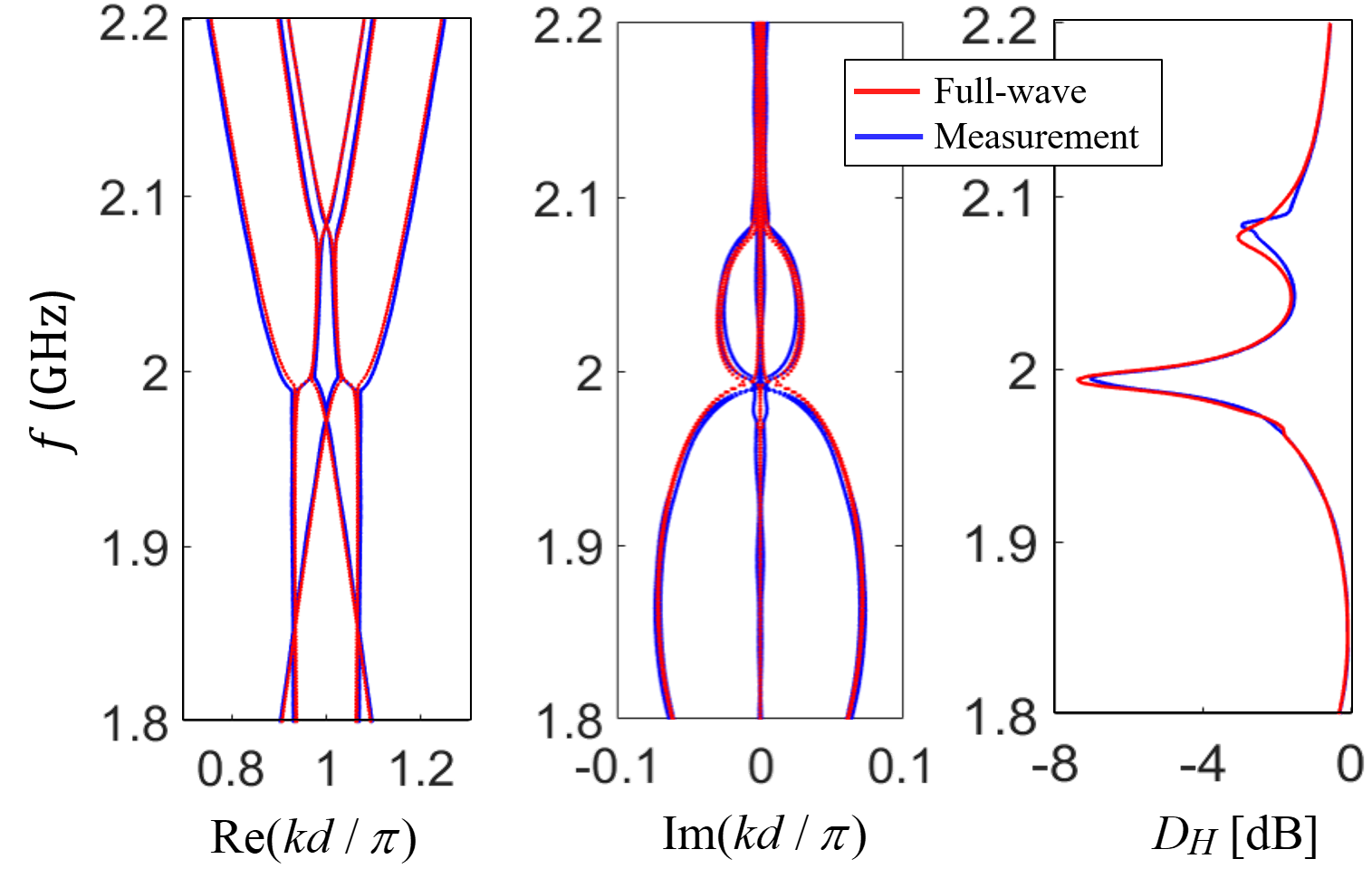}}

\caption{\label{fig:Exp} (a) Fabricated unit cell of the three-way periodic
SIP structure with extra length on both sides to deembed the SMA connectors
effect. (b) Deembedding calibration circuit identical to the extensions
added to the unit-cell. (c) Floquet-Bloch complex-wavenumber dispersion
diagram of the three-way periodic waveguide based on six-port S-parameter
measurements (blue curve), compared to the result obtained from full-wave
simulations of a unit cell S-parameters (red curve). Also, shown in
(c) a comparison in dB between the measured coalescence parameter
and the one obtained from full-wave simulations.}
\end{figure}

We verify the existence of the SIP in the three-way periodic microstrip
shown in Fig. \ref{fig:Proposed_structures}(a) both experimentally
and via full-wave simulations. We used a grounded substrate (Roger5880)
that has\textcolor{black}{{} }substrate loss of $\tan\delta=0.0005$,
whereas the metal layers have conductivity of $4.5\times10^{7}\ $S/m
and thickness of $35\ \mathrm{\mu m}.$ The fabricated unit cell is
shown in Fig. \ref{fig:Exp}(a) including SMA connectors, where we
have added extra extensions of $13$ mm on each side of the unit cell
to deembed the effect of the SMA connectors. Indeed, the SMA connectors
not only add extra length but also introduce high order evanescent
modes due to the discontinuity \cite{apaydin_experimental_2012}.
We fabricated a ``calibration'' circuit with only the two added
extra lengths as shown in Fig. \ref{fig:Exp}(b), to deembed the SMA
effects as follows.

The T-matrices of the two fabricated circuits, the one with the extra
lengths and the calibration circuit, in Figs. \ref{fig:Exp}(a) and
(b) are $\underline{\mathbf{T}}_{A}=\underline{\mathbf{T}}_{R}\underline{\mathbf{T}}_{u}\underline{\mathbf{T}}_{L}$
and $\underline{\mathbf{T}}_{B}=\underline{\mathbf{T}}_{R}\underline{\mathbf{T}}_{L}$,
respectively, where $\underline{\mathbf{T}}_{R}$ and $\underline{\mathbf{T}}_{L}$
are the T-matrices of the extra length and the SMA connectors on the
right and left sides, respectively. Hence, we calculate a new T-matrix
$\underline{\mathbf{T}}_{n}=\underline{\mathbf{T}}_{A}\underline{\mathbf{T}}_{B}^{-1}=\underline{\mathbf{T}}_{R}\underline{\mathbf{T}}_{u}\underline{\mathbf{T}}_{R}^{-1}$
whose eigenvalues are the same of those of the unit-cell T-matrix
$\underline{\mathbf{T}}_{u}$, if $\underline{\mathbf{T}}_{R}$ is
not singular, as shown in \cite{mealy2020general}; a related but
different method is also shown in \cite{apaydin_experimental_2012}.
The transfer matrices $\underline{\mathbf{T}}_{A}$ and $\underline{\mathbf{T}}_{B}$
are obtained by transforming the scattering matrices associated to
the 6-port circuits in Figs. \ref{fig:Exp}(a) and (b).

The scattering matrices are obtained via measurements using a Rohde
\& Schwarz Vector Network Analyzer (VNA) ZVA 67 and also via \textcolor{black}{full-wave
simulations based on the finite element method implemented in CST
Studio Suite}. The measured $6\times6$ S-matrix is obtained through
connecting two ports of the VNA to ports $q$ and $r$ of the unit
cell, while the other four ports are terminated by 50 \textgreek{W}
loads so that we measure a $2\times2$ S-matrix block $\underline{\underline{\mathbf{S}}}(q,r)$.
We change the ports $q$ and $r$ to cover all the combinations of
the 6-ports circuit to construct the $6\times6$ S-matrix from the
obtained $2\times2$ block matrices. Once the T-matrices $\underline{\mathbf{T}}_{A}$
and $\underline{\mathbf{T}}_{B}$ are obtained from the S-parameters,
we calculate the unit-cell complex wavenumbers $k$ from the eigenvalues
of $\underline{\mathbf{T}}_{n}=\underline{\mathbf{T}}_{A}\underline{\mathbf{T}}_{B}^{-1}$
following the method discussed in the previous section. The comparison
between the measured and the simulated dispersion diagrams in Fig.
\ref{fig:Exp}(c) shows a good agreement. \textcolor{black}{As mentioned
previously, conductor, dielectric, and radiation losses slightly affect
the coalescence of the three eigenmodes at the SIP. To quantify this
effect, we calculate the coalescence parameter }$D_{H}$ \textcolor{black}{shown
in Fig. }\ref{fig:Exp}(c)\textcolor{black}{{} using the eigenvectors
of the T-matrix $\underline{\mathbf{T}}_{n}$ obtained from both the
measured and the numerically simulated unit-cell T-matrix. Note that
the eigenvectors of $\underline{\mathbf{T}}_{n}$ and $\underline{\mathbf{T}}_{u}$
are not the same, yet they share the geometrical and algebraic multiplicities
of the eigenvalues, see Ch. 7.2 in \cite{meyer_matrix_2001}, so they
demonstrate similar trends in their coalescence parameters. The dip
at the SIP frequency in the well matched numerically and experimentally
calculated coalescence parameter verifies the existence of the SIP.}

\section{Conclusion}

We have proposed a three-way periodic microstrip waveguide geometry
that exhibits an SIP in its dispersion diagram, without the need of
breaking the system reciprocity. We have provided a theoretical model
to describe the occurrence of the frozen mode regime through the coalescence
parameter that quantifies the coalescence of the eigenvectors and
also via the degeneracy in the complex wavenumber dispersion diagram.
We have demonstrated the occurrence of the frozen mode regime using
both full-wave simulations and scattering parameters measurements
of a six-port unit cell. The SIP in periodic microstrip structures
can serve various applications like distributed amplifiers \cite{Yazdi_Third_2017},
delay lines\cite{Sertel_Harnessing_2019}, pulse generators, and sensors.

\bibliographystyle{IEEEtran}
\bibliography{SIP_CTL_Refs}

\begin{thebibliography}{10}
\providecommand{\url}[1]{#1}
\csname url@samestyle\endcsname
\providecommand{\newblock}{\relax}
\providecommand{\bibinfo}[2]{#2}
\providecommand{\BIBentrySTDinterwordspacing}{\spaceskip=0pt\relax}
\providecommand{\BIBentryALTinterwordstretchfactor}{4}
\providecommand{\BIBentryALTinterwordspacing}{\spaceskip=\fontdimen2\font plus
\BIBentryALTinterwordstretchfactor\fontdimen3\font minus
  \fontdimen4\font\relax}
\providecommand{\BIBforeignlanguage}[2]{{%
\expandafter\ifx\csname l@#1\endcsname\relax
\typeout{** WARNING: IEEEtran.bst: No hyphenation pattern has been}%
\typeout{** loaded for the language `#1'. Using the pattern for}%
\typeout{** the default language instead.}%
\else
\language=\csname l@#1\endcsname
\fi
#2}}
\providecommand{\BIBdecl}{\relax}
\BIBdecl

\bibitem{figotin_oblique_2003}
A.~Figotin and I.~Vitebskiy, ``Oblique frozen modes in periodic layered
  media,'' \emph{Physical Review E}, vol.~68, no.~3, p. 036609, 2003.

\bibitem{mealy2019degeneracy}
T.~Mealy, A.~F. Abdelshafy, and F.~Capolino, ``The degeneracy of the dominant
  mode in rectangular waveguide,'' in \emph{2019 United States National
  Committee of URSI National Radio Science Meeting (USNC-URSI NRSM)}, Boulder,
  CO, USA, 2019, pp. 1-2.

\bibitem{figotin_gigantic_2005}
A.~Figotin and I.~Vitebskiy, ``Gigantic transmission band-edge resonance in
  periodic stacks of anisotropic layers,'' \emph{Physical review E}, vol.~72,
  no.~3, p. 036619, 2005.

\bibitem{nada_theory_2017}
M.~Y. Nada, M.~A. Othman, and F.~Capolino, ``Theory of coupled resonator
  optical waveguides exhibiting high-order exceptional points of degeneracy,''
  \emph{Physical Review B}, vol.~96, no.~18, p. 184304, 2017.

\bibitem{ruter_observation_2010}
C.~E. Rüter, K.~G. Makris, R.~El-Ganainy, D.~N. Christodoulides, M.~Segev, and
  D.~Kip, ``Observation of parity-time symmetry in optics,'' \emph{Nature
  Physics}, vol.~6, no.~3, pp. 192--195, Mar. 2010.

\bibitem{Heiss_PTsymmetry_2017}
J.~Schnabel, H.~Cartarius, J.~Main, G.~Wunner, and W.~D. Heiss,
  ``$\mathcal{PT}$-symmetric waveguide system with evidence of a third-order
  exceptional point,'' \emph{Phys. Rev. A}, vol.~95, p. 053868, May 2017.

\bibitem{Yazdi_Third_2017}
F.~{Yazdi}, M.~A.~K. {Othman}, M.~{Veysi}, F.~{Capolino}, and A.~{Figotin},
  ``Third order modal degeneracy in waveguids: Features and application in
  amplifiers,'' in \emph{2017 USNC-URSI Radio Science Meeting (Joint with AP-S
  Symposium)}, July 2017, pp. 109--110.

\bibitem{Ramezani_Lasing_SIP_2014}
H.~Ramezani, S.~Kalish, I.~Vitebskiy, and T.~Kottos, ``Unidirectional lasing
  emerging from frozen light in nonreciprocal cavities,'' \emph{Physical review
  letters}, vol. 112, no.~4, p. 043904, 2014.

\bibitem{figotin_nonreciprocal_2001}
A.~Figotin and I.~Vitebsky, ``Nonreciprocal magnetic photonic crystals,''
  \emph{Physical Review E}, vol.~63, no.~6, p. 066609, 2001.

\bibitem{figotin_EM_SIP_2003}
A.~Figotin and I.~Vitebskiy, ``Electromagnetic unidirectionality in magnetic
  photonic crystals,'' \emph{Physical Review B}, vol.~67, no.~16, p. 165210,
  2003.

\bibitem{Mumcu_RF_MPC_2005}
G.~{Mumcu}, K.~{Sertel}, J.~L. {Volakis}, I.~{Vitebskiy}, and A.~{Figotin},
  ``Rf propagation in finite thickness unidirectional magnetic photonic
  crystals,'' \emph{IEEE Transactions on Antennas and Propagation}, vol.~53,
  no.~12, pp. 4026--4034, Dec 2005.

\bibitem{Sertel_frozen_2008}
M.~B. {Stephanson}, K.~{Sertel}, and J.~L. {Volakis}, ``Frozen modes in coupled
  microstrip lines printed on ferromagnetic substrates,'' \emph{IEEE Microwave
  and Wireless Components Letters}, vol.~18, no.~5, pp. 305--307, May 2008.

\bibitem{apaydin_experimental_2012}
N.~Apaydin, L.~Zhang, K.~Sertel, and J.~L. Volakis, ``Experimental {Validation}
  of {Frozen} {Modes} {Guided} on {Printed} {Coupled} {Transmission} {Lines},''
  \emph{IEEE Transactions on Microwave Theory and Techniques}, vol.~60, no.~6,
  pp. 1513--1519, Jun. 2012.

\bibitem{Li_Frozen_2017}
H.~Li, I.~Vitebskiy, and T.~Kottos, ``Frozen mode regime in finite periodic
  structures,'' \emph{Phys. Rev. B}, vol.~96, p. 180301, Nov 2017.

\bibitem{Gan_Effects_2019}
Z.~M. Gan, H.~Li, and T.~Kottos, ``Effects of disorder in frozen-mode light,''
  \emph{Optics Letters}, vol.~44, no.~11, pp. 2891--2894, 2019.

\bibitem{Sertel_3way_SIP_2019}
R.~{Almhmadi} and K.~{Sertel}, ``Frozen-light modes in 3-way coupled silicon
  ridge waveguides,'' in \emph{2019 United States National Committee of URSI
  National Radio Science Meeting (USNC-URSI NRSM)}, Jan 2019, pp. 1--2.

\bibitem{Sumetsky_Fiber_2005}
M.~Sumetsky, ``Uniform coil optical resonator and waveguide: transmission
  spectrum, eigenmodes, and dispersion relation,'' \emph{Opt. Express},
  vol.~13, no.~11, pp. 4331--4340, May 2005.

\bibitem{Scheuer_Optical_2011}
J.~Scheuer and M.~Sumetsky, ``Optical-fiber microcoil waveguides and resonators
  and their applications for interferometry and sensing,'' \emph{Laser \&
  Photonics Reviews}, vol.~5, no.~4, pp. 465--478, 2011.

\bibitem{gutman_slow_2012}
N.~Gutman, C.~Martijn~de Sterke, A.~A. Sukhorukov, and L.~C. Botten, ``Slow and
  frozen light in optical waveguides with multiple gratings: {Degenerate} band
  edges and stationary inflection points,'' \emph{Physical Review A}, vol.~85,
  no.~3, p. 033804, Mar. 2012.

\bibitem{wheeler1965transmission}
H.~A. Wheeler, ``Transmission-line properties of parallel strips separated by a
  dielectric sheet,'' \emph{IEEE Transactions on Microwave Theory and
  Techniques}, vol.~13, no.~2, pp. 172--185, 1965.

\bibitem{EASWARAN_Relationship_1993}
V.~Easwaran, V.~Gupta, and M.~Munjal, ``Relationship between the impedance
  matrix and the transfer matrix with specific reference to symmetrical,
  reciprocal and conservative systems,'' \emph{Journal of Sound and Vibration},
  vol. 161, no.~3, pp. 515 -- 525, 1993.

\bibitem{abdelshafy_exceptional_2018}
A.~F. {Abdelshafy}, M.~A.~K. {Othman}, D.~{Oshmarin}, A.~T. {Almutawa}, and
  F.~{Capolino}, ``Exceptional points of degeneracy in periodic coupled
  waveguides and the interplay of gain and radiation loss: Theoretical and
  experimental demonstration,'' \emph{IEEE Transactions on Antennas and
  Propagation}, vol.~67, no.~11, pp. 6909--6923, 2019.

\bibitem{mealy2020general}
T.~Mealy and F.~Capolino, ``General conditions to realize exceptional points of
  degeneracy in two uniform coupled transmission lines,'' \emph{arXiv preprint
  arXiv:2003.04215}, 2020.

\bibitem{meyer_matrix_2001}
C.~D. Meyer, \emph{\BIBforeignlanguage{English}{Matrix analysis and applied
  linear algebra}}.\hskip 1em plus 0.5em minus 0.4em\relax Philadelphia: SIAM:
  Society for Industrial and Applied Mathematics, Feb. 2001.

\bibitem{Sertel_Harnessing_2019}
B.~{Paul}, N.~K. {Nahar}, and K.~{Sertel}, ``Harnessing the frozen-mode in
  coupled silicon ridge waveguides for true time delay applications,'' in
  \emph{2019 International Conference on Electromagnetics in Advanced
  Applications (ICEAA)}, Granada, Spain, 2019, pp. 0552--0552.

\end{thebibliography}

\end{document}